\documentclass[a4paper,12pt]{article}
\usepackage{amssymb}

\newtheorem{theorem}{Theorem}
\newtheorem{corollary}{Corollary}
\newtheorem{claim}{Claim}
\newtheorem{remark}{Remark}

\begin{document}
\author{Siham BEKKAI\thanks{%
e-mail address: siham.bekkai@gmail.com} \\
\centerline{\small\textit{USTHB, Faculty of Mathematics, PO Box 32 El-Alia
Bab Ezzouar}}
\\ \centerline{\small\textit{16111 Algiers, Algeria}}}

\title{Minimum Degree, Independence Number and Pseudo $[2,b]$-Factors in Graphs}
\date{}
\maketitle{}

\begin{abstract}
A pseudo $[2,b]$-factor of a graph $G$ is a spanning subgraph in
which each component $C$ on at least three vertices verifies $2\leq
d_C(x)\leq b$, for every vertex $x$ in $C$. The main contibution of this
paper, is to give an upper bound to the number of components that
are edges or vertices in a pseudo $[2,b]$-factor of a graph $G$.
Given an integer $b\geq4$, we show that a graph $G$ with minimum
degree $\delta$, independence number $\alpha>\frac{b(\delta-1)}{2}$
and without isolated vertices possesses a pseudo $[2,b]$-factor with
at most $\alpha-\lfloor\frac{b}{2}(\delta-1)\rfloor$ edges or
vertices. This bound is sharp.

\end{abstract}


\section{Introduction}
Throughout this paper, graphs are assumed to be finite and simple.
For unexplained concepts and notations, the reader could refer to
\cite{bon}.

Given a graph $G$, we let $V(G)$ be its vertex set, $E(G)$ its edge
set and $n$ its order. The neighborhood of a vertex $x$ in $G$ is
denoted by $N_G(x)$ and defined to be the set of vertices of $G$
adjacent to $x$; the cardinality of this set is called the degree of
$x$ in $G$. For convenience, we denote by $d(x)$ the degree of a vertex
$x$ in $G$; by $\delta$ the minimum degree of $G$ and by $\alpha$ its independence number.
However, if $H$ is a subgraph of $G$ then we write $d_H(x)$;
$\delta_H$ and $\alpha(H)$ respectively for the degree of $x$ in
$H$; the minimum degree and the independence number of $H$. We
denote by $d_G(x,y)$ the distance between $x$ and $y$ in the graph
$G$.

A factor of $G$ is a spanning subgraph of $G$, that is a subgraph
obtained by edge deletions only. If $S$ is the set of deleted edges,
then this subgraph is denoted $G-S$.
If $H$ is a subgraph of $G$, then $G-H$ stands for the subgraph induced by
$V(G)-V(H)$ in $G$. By starting with a disjoint union of two graphs $G_1$ and $G_2$ and
adding edges joining every vertex of $G_1$ to every vertex of $G_2$, we
obtain the join of $G_1$ and $G_2$, denoted $G_1+G_2$. For a positive
integer $p$, the graph $pG$ consists of $p$ vertex-disjoint copies
of $G$. In all what follows, we use disjoint to stand for vertex-disjoint.

In \cite{bek}, we defined a pseudo 2-factor of a graph $G$
to be a factor each component of which is a cycle, an edge or a
vertex. It can also be seen as a graph partition by a family of vertices, edges and cycles.
Graph partition problems have been studied in lots of papers. They consist in partitioning the vertex set of $G$
by disjoint subgraphs chosen to have some specific properties. In \cite{EnomotoSurvey}, Enomoto listed a variety
of results dealing with partitions into paths and cycles. The emphasis is generally on the existence of
a given partition however, in our study of pseudo-factors, we take interest in the number of components that
are edges or vertices in a pseudo-factor of $G$. In \cite{bek}, we proved that every graph with minimum degree
$\delta\geq1$ and independence number $\alpha\geq\delta$ possesses a
pseudo 2-factor with at most $\alpha-\delta+1$ edges or vertices and
that this bound is best possible. Motivated by the desire to know what happens in general cases, we define a
\emph{pseudo $[a,b]$-factor} (where $a$ and $b$ are two integers such that $b\geq a\geq 2$) as a factor of $G$
in which each component $C$ on at least three vertices verifies
$a\leq d_C(x)\leq b$, for every $x\in C$. Clearly, a pseudo
$[a,b]$-factor with no component that is an edge or a vertex is
nothing but an $[a,b]$-factor. Surveys on factors and specifically $[a,b]$-factors
and connected factors can be found in \cite{PlummerSurveyFactors, KouiderVestergaard}.
In the present work, we study pseudo $[2,b]$-factors, we consider the case $b\geq4$ and obtain an upper
bound (in function of $\delta$, $\alpha$ and $b$) for the number of
components that are edges or vertices in a pseudo $[2,b]$-factor of
$G$. Note that, from a result by Kouider and Lonc
(\cite{KouiderLonc}), we deduce that if
$\alpha\leq\frac{b(\delta-1)}{2}$ then $G$ has a $[2,b]$-factor.
Laying down the condition $\alpha>\frac{b(\delta-1)}{2}$, the main
result of this paper reads as follows:

\begin{theorem}\label{thm1}
Let $b$ be an integer such that $b\geq4$ and $G$ a graph of minimum
degree $\delta\geq1$ and independence number $\alpha$ with
$\alpha>\frac{b(\delta-1)}{2}$. Then $G$ possesses a pseudo
$[2,b]$-factor with at most
$\alpha-\lfloor\frac{b}{2}(\delta-1)\rfloor$ components that are
edges or vertices.
\end{theorem}

The bound given in Theorem \ref{thm1} is best possible. Indeed, let
$b$ be an integer such that $b\geq4$ and let $H$ be a nonempty set
of vertices. The graph $G=H+pK_2$, where $p>\frac{b}{2}|H|$, has
minimum degree $\delta=|H|+1$ and independence number $\alpha=p$. We
can easily verify that $G$ possesses a pseudo $[2,b]$-factor with
$\alpha-\lfloor\frac{b}{2}(\delta-1)\rfloor$ edges and we can not do
better. Also, a simple example reaching the bound of Theorem
\ref{thm1}, is a graph $G$ obtained by taking a graph $H$ on $n$
vertices in which every vertex is of degree between 2 and $b$
($b\geq4$), then taking $n$ additional independent vertices and
joining exactly one isolated vertex to exactly one vertex of $H$.
The graph $G$ has minimum degree $\delta=1$, independence number
$\alpha=n$ and can be partitioned into one component that is $H$ and
$n=\alpha-\lfloor\frac{b}{2}(\delta-1)\rfloor$ vertices (or simply
$n$ edges) and we can not do better.

Combining Theorem \ref{thm1} with the results of \cite{bek}
and \cite{KouiderLonc}, we obtain

\begin{corollary}
Let $b\geq2$ be an integer such that $b\neq3$. Let $G$ be a graph of
minimum degree $\delta$ and independence number $\alpha$ and without
isolated vertices. Then $G$ possesses a pseudo $[2,b]$-factor with
at most $\max(0,\alpha-\lfloor\frac{b}{2}(\delta-1)\rfloor)$ edges
or vertices.
\end{corollary}

\section{Independence number, minimum degree and pseudo $[2,b]$-factors}

First of all, we put aside the case $\delta=1$ for which we know
that we have in $G$ a pseudo $[2,b]$-factor with at most $\alpha$
edges or vertices. Indeed, if we regard a cycle as a component each
vertex of which is of degree between $2$ and $b$, then we know that
any graph $G$ can be covered by at most $\alpha$ cycles, edges or
vertices (see for instance \cite{posa}). So the bound
$\alpha-\lfloor\frac{b}{2}(\delta-1)\rfloor$ holds for $\delta=1$.

From now on, we assume that $G$ has minimum degree $\delta\geq2$.
Let $F$ be a subgraph of $G$ such that $2\leq d_F(x)\leq b$ for all
$x\in V(F)$. For the sake of simplifying the writing, such a
subgraph $F$ will be called a $[2,b]$-subgraph of $G$. Denote by $D$
a smallest component of $G-F$, set $W=G-(D\cup F)$ and choose $F$ in
such a manner that:

$(a)$ $\alpha(G-F)$ is as small as possible;

$(b)$ subject to $(a)$, the number of vertices of $D$ is as small as
possible;

$(c)$ subject to $(a)$ and $(b)$, the number of vertices in $F$ is
as small as possible.

Notice that a subgraph $F$ satisfying the conditions above exists
since $\delta\geq2$. Indeed, let us consider a longest path in $G$
and let $u$ be one of its endpoints. Let $v$ be the farthest
neighbor of $u$ on this path and $P_{uv}$ the segment of $P$ joining
$u$ and $v$. The cycle $C$ formed by the path $P_{uv}$ and the edge $uv$ contains $u$ and all its
neighbors so $\alpha(G-C)<\alpha$. Hence $F$ is not empty.

We shall show the following theorem which yields Theorem \ref{thm1}:

\begin{theorem}\label{thm2}
Let $b$ be an integer such that $b\geq4$. Let $G$ be a graph of
minimum degree $\delta\geq2$ and independence number $\alpha$ such
that $\alpha>\frac{b(\delta-1)}{2}$. Then there exists a pseudo
$[2,b]$-factor of $G$ such that $F$ is the $[2,b]$-subgraph of this
pseudo $[2,b]$-factor and $F$ gives
$\alpha(G-F)\leq\alpha-\lfloor\frac{b}{2}(\delta-1)\rfloor$.

\end{theorem}

\textbf{Proof of Theorem\ref{thm2}.} Let $F$ be a $[2,b]$-subgraph
of $G$ satisfying the conditions $(a)$, $(b)$ and $(c)$. Denote by
$u_1,...,u_m$ ($m\geq1$) the neighbors of $D$ on $F$ and by $P_{ij}$
a path with internal vertices in $D$ joining two vertices $u_i$ and
$u_j$ with $1\leq i,j\leq m$ and $i\neq j$. The proof of Theorem
\ref{thm2} will be divided into several claims. The following one
which will be intensively used reminds Lemma 1 in
\cite{bek}.

\begin{claim}\label{claim1}
Let $F'$ be a $[2,b]$-subgraph of $G$ which contains the neighbors
of $D$ in $F$ and at least one vertex of $D$. Setting $W'=G-(F'\cup
D)$, we have $\alpha(W')>\alpha(W)$.
\end{claim}

\emph{Proof of Claim \ref{claim1}.} Set $D'=D-F'$.
\\ (1) If $D'=\emptyset$ then by the choice of $F$, we have
$\alpha(G-F)\leq\alpha(G-F')$. But
\\ $\alpha(G-F)=\alpha(W)+\alpha(D)\geq\alpha(W)+1$ and $\alpha(G-F')=\alpha(W')$, so
$\alpha(W)<\alpha(W')$.
\\ (2) If $D'\neq\emptyset$ then $F'$ gives a component $D'$ smaller than $D$, so again by the choice of $F$, we have
$\alpha(W)+\alpha(D)=\alpha(G-F)<\alpha(G-F')=\alpha(W')+\alpha(D')$.
But as $\alpha(D')\leq\alpha(D)$ then we obtain
$\alpha(W)<\alpha(W')$. $\Box$

\vspace{3mm} In the next claims, we try to learn more about the
degrees in $F$ of its vertices.

\begin{claim}\label{claim2}
For every $i$, $1\leq i\leq m$, we have
$N_F(u_i)\cap\{u_1,\ldots,u_m\}=\emptyset$.
\end{claim}

\emph{Proof of Claim \ref{claim2}.} Suppose that for some $i$,
$N_F(u_i)\cap\{u_1,...,u_m\}\neq\emptyset$, then there exists a
vertex $u_j$ ($1\leq j\leq m$ and $j\neq i$) such that $u_iu_j\in
E(F)$. Put $e=u_iu_j$, then $(F-e)\cup P_{ij}$ is a
$[2,b]$-subgraph. Indeed, none of the vertices of $F$ changes its
degree in $(F-e)\cup P_{ij}$ and the internal vertices of $P_{ij}$
are of degree 2. So taking $F'=(F-e)\cup P_{ij}$ in Claim
\ref{claim1} we obtain $\alpha(W)>\alpha(W)$, which is absurd.
$\Box$

\begin{claim}\label{claim3}
$d_F(u_i)\leq b-1$ for at most one vertex $u_i$, $i=1,...,m$.

\end{claim}

\emph{Proof of Claim \ref{claim3}.} Suppose to the contrary that
there exist at least two distinct vertices $u_k$ and $u_l$ such that
$d_F(u_k)\leq b-1$ and $d_F(u_l)\leq b-1$. Then taking $F'=F\cup
P_{kl}$ in Claim \ref{claim1} (notice that in $F'$, $d_F(u_k)$ and
$d_F(u_l)$ are at most $b$, and the internal vertices of $P_{kl}$
are of degree 2 in $F'$ so $F'$ is a $[2,b]$-subgraph of $G$), we
obtain $\alpha(W)<\alpha(W)$ which is absurd. $\Box$

\vspace{3mm}

Let $S$ be the set of vertices $x$ in $\cup_{i=1}^m N_F(u_i)$ such
that $x$ is a common neighbor of at least two vertices in
$\{u_1,...,u_m\}$. We have:

\begin{claim}\label{claim+}
\begin{enumerate}
\item $d_F(x)\leq3$ for every $x\in S$.
\item If $S$ contains a vertex $x$ such that $d_F(x)=3$, then

\begin{enumerate}
\item For every $k, 1\leq k\leq m$, we have $d_F(u_k)=b$.
\item For every $y\in \cup_{i=1}^m N_F(u_i)-\{x\}$ we have $d_F(y)=2$.
\end{enumerate}
\end{enumerate}
\end{claim}

\emph{Proof of Claim \ref{claim+}.}
\begin{enumerate}
\item Suppose that $d_F(x)\geq4$ for some $x\in S$. By definition, $x$
is the neighbor in $F$ of at least two vertices say $u_i$ and $u_j$
with $1\leq i, j\leq m, i\neq j$. Put $e=xu_i$ and $e'=xu_j$. Then
in $F'=(F-e-e')\cup P_{ij}$ only $x$ changes its degree but it
remains at least 2. So $F'$ is a $[2,b]$-subgraph which leads to a
contradiction by Claim \ref{claim1}.

\item Let $x$ be in $N_F(u_i)\cap N_F(u_j)$ ($1\leq i,j\leq m$ and $i\neq j$) such that $d_F(x)=3$.
Suppose that there exists $u_k$ (which will be the only one by Claim
\ref{claim3}) such that $d_F(u_k)\leq b-1$, we can always assume
that $k\neq i$. Then taking $F'=(F-e)\cup P_{ik}$, where $e=xu_i$,
in Claim \ref{claim1} gives a contradiction.
\\ Furthermore, if we suppose that there exists $y\in
N_F(u_k)-\{x\}$, with $1\leq k\leq m$ (we can suppose without loss
of generality that $k\neq i$) such that $d_F(y)\geq3$. Then setting
$e=xu_i$, $e'=yu_k$ and taking $F'=(F-e-e')\cup P_{ik}$ in Claim
\ref{claim1} gives a contradiction. Notice that $F'$ is a
$[2,b]$-subgraph: indeed, only $x$ and $y$ lose 1 in their degree
but they remain of degree at least 2 in $F'$ and the internal
vertices of $P_{ik}$ are of degree 2 in $F'$.$\Box$
\end{enumerate}

\vspace{3mm}

Claim \ref{claim+} implies that $S$ is an independent set in $F$ and
we will deduce later that it is also independent in $G$. But before
that, we take a look at the neighbors of $\{u_1,...,u_m\}$ which are
not in $S$. For each $u_i$ ($i=1,...,m$), set
$N^*_F(u_i)=\{x\in N_F(u_i); x\notin S\}$.

\begin{claim}\label{claim4}
\begin{enumerate}
\item If there exist vertices $x$ in $\cup_{i=1}^m N^*_F(u_i)$ such that
$d_F(x)\geq3$, then these vertices are in the neighborhood of a same
$u_k$, $1\leq k\leq m$.
\item If there exist $k, 1\leq k\leq m$, such that $d_F(u_k)\leq b-1$
and $x\in \cup_{i=1}^m N_F(u_i)$ such that $d_F(x)\geq3$, then $x\in
N^*_F(u_k)$.
\end{enumerate}
\end{claim}

\emph{Proof of Claim \ref{claim4}.}
\begin{enumerate}
\item Suppose that there exist $x\in N^*_F(u_k)$ such that $d_F(x)\geq3$,
$x'\in N^*_F(u_j)$ such that $d_F(x')\geq3$ and $1\leq j,
k\leq m, j\neq k$. Then the subgraph $(F-e-e')\cup P_{kj}$, where $e=u_kx$
and $e'=u_jx'$ is a $[2,b]$-subgraph of $G$. Taking $F'=(F-e-e')\cup
P_{kj}$ in Claim \ref{claim1}, we obtain a contradiction.
\item Suppose that there exist $k, 1\leq k\leq m$, such that $d_F(u_k)\leq b-1$ and
$x\in \cup_{i=1}^m N_F(u_i)$ with $d_F(x)\geq3$. By Claim
\ref{claim+}(2), $x\notin S$. Suppose that $x\in N^*_F(u_i)$, with
$i\neq k$. Notice that the fact that $d_F(u_k)\leq b-1$ forces
$d_F(u_i)$, by Claim \ref{claim3}, to be equal to $b$. Taking
$F'=(F-e)\cup P_{ik}$, where $e=xu_i$, in Claim \ref{claim1}, we
obtain $\alpha(W)>\alpha(W)$ which is absurd. $\Box$
\end{enumerate}

\vspace{3mm}

Looking more closely at the structure of $D$, we can say
more about the degrees of the vertices in $\cup_{i=1}^m N_F[u_i]$,
where $N_F[u_i]=N_F(u_i)\cup\{u_i\}$ is the closed neighborhood of
$u_i$. First, we remark that $D$ has minimum degree at most 1.

\begin{remark}\label{rmk1}
$\delta_D\leq1$.
\end{remark}
\emph{Proof.} Suppose, by contradiction, that $\delta_D\geq2$ then
taking a longest path in $D$ provides a cycle $C$ which verifies
$\alpha(D-C)<\alpha(D)$. Put $F'=F\cup C$, then $F'$ is a
$[2,b]$-subgraph of $G$. Moreover,
$\alpha(G-F')=\alpha(D-C)+\alpha(W)<\alpha(D)+\alpha(W)=\alpha(G-F)$
and this contradicts the choice of $F$.$\Box$

\vspace{3mm}

Two cases are to consider, the case where $D$ is a tree (a single
vertex is a trivial tree) and the case where $D$ contains a cycle.
The following claim deals with this latter case.

\begin{claim}\label{claimDcycle}
Suppose that $D$ contains a cycle. Then
\begin{enumerate}
\item $d_F(x)=2$ for all $x\in \cup_{i=1}^m N_F(u_i)$.
\item $d_F(u_i)=b$ for all $i, 1\leq i\leq m$.
\end{enumerate}
\end{claim}
\newpage
\emph{Proof of Claim \ref{claimDcycle}.}
\begin{enumerate}
\item Suppose that there exists a vertex $x\in N_F(u_i)$ such that
$d_F(x)\geq3$ and let $Q$ be an edge or a path with internal
vertices in $D-C$ joining $u_i$ and $C$. Then taking $F'=(F-e)\cup
Q\cup C$, where $e=xu_i$, in Claim 1 gives a contradiction. Notice
that $d_{F'}(x)\geq2$ and that $u_i$ does not change its degree (nor
do the other vertices of $F$) then $F'$ is a $[2,b]$-subgraph of
$G$.
\item Suppose that
$d_F(u_k)\leq b-1$ for some $k, 1\leq k\leq m$ and let $Q$ be an
edge or a path with internal vertices in $D-C$ joining $u_k$ and
$C$. Then, taking $F'=F\cup Q\cup C$ in Claim 1 gives a
contradiction. $\Box$
\end{enumerate}

\vspace{3mm}

If $D$ is a tree and $\delta_D\neq0$, then $D$ has at least two leaves,
say $x_0$ and $y_0$. We relabel $u_1,...,u_{m_1}$, with $m_1\leq m$,
the vertices in $N_F(x_0)\cup N_F(y_0)$.

\begin{claim}\label{claim7}
Suppose that $D$ is a tree and that there exist two vertices $x_0$
and $y_0$ in $D$ with $d_D(x_0)=d_D(y_0)=1$ such that
$N_F(x_0)=N_F(y_0)$. Then
\begin{enumerate}

\item For all $k, 1\leq k\leq m_1$, $d_F(u_k)\geq b-1$.
\item If there exists a vertex $x\in \cup_{i=1}^{m_1}
N_F(u_i)$ such that $d_F(x)\geq3$ then it is the only one.
\item If there exists a vertex $u_k$ (with $1\leq k\leq m_1$) such that
$d_F(u_k)=b-1$ then for every vertex $x\in \cup_{i=1}^{m_1}N_F(u_i)$
we have $d_F(x)=2$.
\end{enumerate}
\end{claim}

\emph{Proof of Claim \ref{claim7}.} Let $P$ be a path in $D$ joining
$x_0$ to $y_0$.
\begin{enumerate}

\item If there exists a vertex $u_i$ ($1\leq i\leq m_1$) such that
$d_F(u_i)\leq b-2$. Then taking $F'=F\cup u_ix_0Py_0u_i$ in Claim
\ref{claim1} gives a contradiction.

\item Suppose that there exists a vertex $x\in \cup_{i=1}^{m_1}N_F(u_i)$ such that $d_F(x)\geq3$.
If $x\in S$ then Claim \ref{claim+} gives what desired. If $x\notin
S$, then $x\in N^*_F(u_k)$ for some $1\leq k\leq m_1$. Suppose that
there exists $y\in\cup_{i=1}^{m_1}N_F(u_i)$ such that $d_F(y)\geq3$
so $y\in N^*_F(u_j)$, for some $1\leq j\leq m_1$. By Claim
\ref{claim4}(1), $j=k$. Taking $F'=(F-e-e')\cup u_kx_0Py_0u_k$,
where $e=xu_k$ and $e'=yu_k$, in Claim \ref{claim1}, we obtain a
contradiction.

\item Finally, suppose that there exist a vertex $u_k$ ($1\leq k\leq m_1$) such that $d_F(u_k)=b-1$
and a vertex $x\in \cup_{i=1}^{m_1}N_F(u_i)$ such that
$d_F(x)\geq3$. By Claim \ref{claim4}(2), $x\in N^*_F(u_k)$. Put
$e=xu_k$. Then taking $F'=(F-e)\cup u_kx_0Py_0$ in Claim
\ref{claim1} gives a contradiction. $\Box$\end{enumerate}

\vspace{3mm} A path $I$ in $F$ with $V(I)\subset V(F)$, $E(I)\subset
E(F)$ and such that every internal vertex $x$ of $I$ has $d_F(x)=2$
is called an interval (or a segment) of $F$. We say that two
disjoint intervals $I^{(1)}$ and $I^{(^2)}$ in $F$ are
\emph{path-independent} if there exists no path internally disjoint
from $F\cup D$ joining a vertex in $I^{(1)}$ to a vertex in
$I^{(2)}$. We say that $t$ intervals $I^{(1)}, I^{(2)},
\ldots,I^{(t)}$ ($t\geq2$) in $F$ are path-independent if they are
pairwise path-independent. The following claim will be very useful.
It is a shorter version of Lemma 2 in \cite{bek} with a
short proof.

\begin{claim}\label{claim*}
Let $I^{(1)}, I^{(2)},\ldots,I^{(t)}$ ($t\geq2$) be $t$
disjoint intervals in $F$, containing no neighbor of $D$ and
such that $\alpha(W\cup I^{(i)})=\alpha(W)$ for every $i=1,...,t$.
If $I^{(1)}, I^{(2)},\ldots,I^{(t)}$ are path-independent, then
$\alpha(D\cup W\cup I^{(1)}\cup I^{(2)}\cup\ldots\cup
I^{(t)})=\alpha(W\cup D)$.
\end{claim}

\emph{Proof of Claim \ref{claim*}.}

Let $W_i$ be the union of components of $W$ with neighbors in
$I^{(i)}$ ($i=1,\ldots,t$). By hypothesis, the intervals $I^{(i)}$
are pairwise path-independent so $W_i\cap W_j=\emptyset$, for all
$1\leq i,j\leq t, i\neq j$. Hence $W-\cup_{i=1}^t W_i,W_1\cup
I^{(1)},\ldots,W_t\cup I^{(t)}$ form a partition of $W\cup
I^{(1)}\cup\ldots\cup I^{(t)}$ and it follows that $\alpha(W\cup
I^{(1)}\cup\ldots\cup I^{(t)})=\alpha(W_1\cup I^{(1)})+\cdots
+\alpha(W_t\cup I^{(t)})+\alpha(W-\cup_{i=1}^t W_i)$. On the other
hand, as $\alpha(W\cup I^{(i)})=\alpha(W)$, for every $i=1,\ldots,t$
then $\alpha(W_i\cup I^{(i)})=\alpha(W_i)$. This yields
$\alpha(W\cup I^{(1)}\cup\ldots\cup
I^{(t)})=\sum_{i=1}^t\alpha(W_i)+ \alpha(W-\cup_{i=1}^t
W_i)=\alpha(W)$. We finally get $\alpha(W\cup D\cup I^{(1)}\cup
\ldots\cup I^{(t)})=\alpha(W\cup D)$ because the intervals $I^{(i)}$
(with $i=1,\ldots,t$) do contain no neighbor of $D$. $\Box$

\vspace{3mm}

Let $s$ be the vertex of $S$ (if it exists) such that $d_F(s)=3$. We
put $s$ aside before applying the procedure described hereafter.
Provided always that $s$ exists, we set $N'_F(u_i)=N_F(u_i)-\{s\}$
if $s\in N_F(u_i)$ and $N'_F(u_i)=N_F(u_i)$ otherwise ($i=1,...,m$).

For $u_k$, $1\leq k\leq m$, denote by $x^k_{i}$
($i=1,...,|N'_F(u_k)|$) its neighbors that belong to $N'_F(u_k)$.
Using this notation, we can have $x^k_i=x^l_j$, for some $1\leq
i\leq|N'_F(u_k)|$ and $1\leq j\leq|N'_F(u_l)|$ ($1\leq k,l\leq m, k\neq l$), in case $x^k_i\in
N'_F(u_k)\cap N'_F(u_l)\subset S$.

From now on, let $m'=m_1$ if $D$ is a tree with at least two leaves
and $m'=m$ otherwise. We choose the sense $u_k\rightarrow x^k_i$, as
a sense of "orientation". Let $u_k$ ($1\leq k\leq m'$) be such that
$d_F(u_k)=b$ and all its neighbors that are in $N'_F(u_k)$ are of
degree 2 in $F$. Starting at $x^k_1$
and following the chosen orientation we go over from a vertex to its
neighbor until meeting a vertex which we call $y^k_1$, such that
$d_F(y^{k,+}_{1})\geq3$ or $y^{k,+}_1=u_j$ for some $j, 1\leq j\leq
m$ (where $y^{k,+}_1$ is the successor of $y^{k}_1$ following the chosen orientation).
This gives an interval $x^k_1...y^k_1$ which we denote by $P_1^k=[x^k_1,y^k_1]_F$.

We repeat the process using the other neighbors of $u_k$ that are in
$N'_F(u_k)$. At the p$^{th}$ step, we consider a vertex $x^k_p\in
X_p= N'_F(u_k)-(\cup_{i=1}^{p-1} V(P^k_i))$ and construct a path
$P^k_p=x_p^k...y_p^k$ containing $x^k_p$ and such that
$d_F(y^{k,+}_p)\geq3$ or $y^{k,+}_p=u_j$ for some $j, 1\leq j\leq
m$. When $X_r$ becomes empty at the r$^{th}$ step ($r\geq p$), then
we consider another vertex $u_l$ ($l\neq k$). We choose as long as
possible, $u_l$ such that $d_F(u_l)=b$ and its neighborhood that are
in $N'_F(u_l)$ are all of degree 2 in $F$. We Choose a vertex
 in $N'_F(u_l)-\cup_{i=1}^{r-1} V(P^k_i)$, and we do the same
construction, until the vertices in $N'_F(u_l)$ are all in
$(\cup_{i=1}^{r-1} V(P^k_i))\cup(\cup_{i=1}^{r'-1} V(P^l_i))$.
Denote by $\mathfrak{P}$ the set of paths obtained so far. When it
is no more possible to choose a vertex $u_p$, $1\leq p\leq m'$, such
that $d_F(u_p)=b$ and with all its neighbors that are in $N'_F(u_p)$
having degree 2 in $F$, then we take the vertex $u_q$ of degree at
most $b-1$ or having in its neighborhood $N'_F(u_q)$ vertices of
degree at least $3$ in $F$. Notice that $u_q$ exists only if $s$
does not (see Claim \ref{claim+}(2)) and if both a vertex $u_q$ of
degree at most $b-1$ (which would be the only one by Claim
\ref{claim3}) and vertices $x_i^j$ of degree at least $3$ exist,
then these vertices are in the neighborhood of $u_q$ (see Claim
\ref{claim4}). Put $N_q=\{x^q_i\in N'_F(u_q)-V(\mathfrak{P})$ such
that $d_F(x^q_i)=2\}$. Starting at a vertex $x^q_i\in N_q$, we
repeat the construction described above until $N_q$ becomes empty.
We update the set $\mathfrak{P}$ at each step.

By construction all the vertices of $P^k_i$ are of degree 2 in $F$
so $V(P^k_i)\cap V(P^l_j)=\emptyset$ for every couple $P^k_i, P^l_j$
of paths in $\mathfrak{P}$ (they are disjoint), moreover no
vertex in $P^k_i$ is adjacent in $F$ to a vertex in $P^l_j$, for all
$P^k_i, P^l_j$ in $\mathfrak{P}$.

We divide the set $\mathfrak{P}$ into three subsets, each containing
the paths $P^k_i=[x^k_i,y^k_i]_F$ of Type 1, Type 2 or Type 3,
defined as follows:

\begin{description}
\item[Type 1] If $y^{k,+}_i=u_j$, with $j\neq k$.
\item[Type 2] If $y^{k,+}_i=u_j$, with $j=k$.
\item[Type 3] If $y^{k,+}_i\neq u_j$ for every $j$, $1\leq j\leq m$.

\end{description}

For technical reasons, in case $D$ is a trivial tree or a tree
having no couple of leaves with the same neighborhood in $F$, we
stop the procedure described above when it remains no vertex $u_p$
($1\leq p\leq m'$) such that $d_F(u_p)=b$, or when the remaining
vertex $u_p$ ($1\leq p\leq m'$) has in its neighborhood $N'_F(u_p)$
a vertex of degree at least 3. We consider $\mathfrak{Q}$ the subset
of $\mathfrak{P}$, of paths obtained till then. Let
$\mathfrak{P_1}=\mathfrak{Q}$ in this case and
$\mathfrak{P_1}=\mathfrak{P}$ in the others.

We show in what follows that the addition of a path of
$\mathfrak{P_1}$ to $W\cup D$ augments $\alpha(W\cup D)$ by at least
1.

\begin{claim}\label{claim8}
For each $P^k_i\in\mathfrak{P_1}$, we have $\alpha(W\cup D\cup
P^k_i)>\alpha(W\cup D)$.
\end{claim}

\emph{Proof of Claim \ref{claim8}.} Let $P^k_i$ be a path in
$\mathfrak{P_1}$.
\begin{enumerate}
\item If $D$ contains a cycle, then taking $F'=F-P^k_i$
gives what desired. Indeed, in this case all the vertices $u_i$ are
of degree $b$ (by Claim \ref{claimDcycle}), as $b\geq4$ then after
the deletion of $P^k_i$, the degree of the vertices $u_i$ ($1\leq
i\leq m'$) remains at least $2$. Moreover, by construction of
$P^k_i$, the degree of no vertex in $F$ becomes smaller than $2$,
after deletion of $P^k_i$. So $F'$ is a $[2,b]$-subgraph of $G$.
$F'$ contradicts Condition $(c)$ in the choice of $F$ (because
$|V(F')|<|V(F)|$) so $\alpha(G-F')>\alpha(G-F)$ which yields
$\alpha(W\cup D\cup P^k_i)>\alpha(W\cup D)$.
\item  If $D$ is a tree possessing two
vertices $x_0$ and $y_0$ of degree $1$ in $D$, having the same
neighborhood in $F$ ($N_F(x_0)=N_F(y_0)$). Then if $P^k_i$ is of
Type 1 or 3, then we reason as in (1) and we obtain what desired. If
$P^k_i$ is of Type 2, then (1) is no more efficient if
$d_F(u_k)=b-1$ (because the degree of $u_k$ may become smaller than
2 when $P^k_i$ is deleted). So we take $F'=(F-P^k_i)\cup
u_kx_0Py_0u_k$, where $P$ is a path with internal vertices in $D$ joining $x_0$ to $y_0$.
The subgraph $F'$ is a $[2,b]$-subgraph of $G$ (we have
$d_F(u_k)=d_{F'}(u_k)$) which gives by Claim \ref{claim1}, what
desired.
\item In the other cases, as $b\geq4$ and by the choice of the subset $\mathfrak{P_1}$, the deletion of
any path $P^k_i\in\mathfrak{P_1}$, gives a $[2,b]$-subgraph.
Reasoning as in (1), we get what desired. $\Box$
\end{enumerate}

\vspace{3mm}

Notice that as $D$ is independent from $P^k_i$ (by construction) and
from $W$ then $\alpha(W\cup D\cup P^k_i)=\alpha(W\cup
P^k_i)+\alpha(D)$. Hence the conclusion in Claim \ref{claim8} is
equivalent to $\alpha(W\cup P^k_i)>\alpha(W)$. For each path
$P^k_i=[x^k_i,y^k_i]_F$ in $\mathfrak{P_1}$ and following the chosen
orientation, let $v^k_i$ be the first vertex of $P^k_i$ such that
$\alpha(W\cup[x^k_i,v^k_i]_F)>\alpha(W)$. Notice that $v^k_i$ is
well defined by Claim \ref{claim8}. Denote by $P^{'k}_{i}$ the interval
$[x^k_i,v^k_i]_F$ of $P^k_i$ and by $\mathfrak{P'}$ the set of the
intervals $P^{'k}_{i}$. In what follows, we take interest in the
path-independence of the intervals of $\mathfrak{P'}$.

\begin{claim}\label{claim9}
Let $P^{'k}_i$ and $P^{'l}_j$ be two distinct intervals
$[x^k_i,v^k_i]_F$ and $[x^l_j,v^l_j]_F$ in $\mathfrak{P'}$ such that
$1\leq k, l\leq m'$, $k\neq l$. Then, $P^{'k}_i$ and $P^{'l}_j$ are
path-independent.

\end{claim}

\emph{Proof of Claim \ref{claim9}.} By way of contradiction, suppose
that there exist two vertices $a^k_i\in P^{'k}_i$ and $a^l_j\in
P^{'l}_j$ such that $a^k_i$ and $a^l_j$ are joined by $Q$ which is
an edge in $G$ or a path with internal vertices in $W$. Choose
$a^k_i$ and $a^l_j$ so as to minimize the sum
$d_F(x^k_i,a^k_i)+d_F(x^l_j,a^l_j)$. Recall that by construction
$xy\notin E(F)$ for every $x\in P^k_i$ and $y\in P^l_j$.
\\ The segments $[x^k_i,a^k_i[_F$ and
$[x^l_j,a^l_j[_F$ verify the hypothesis of Claim \ref{claim*}.
Indeed, by the choice of $a^k_i$ and $a^l_j$, they are
path-independent. Furthermore, as
$[x^k_i,a^k_i[_F\subset[x^k_i,v^k_i[_F$;
$[x^l_j,a^l_j[_F\subset[x^l_j,v^l_j[_F$ and by the choice of $v^k_i$
and $v^l_j$, we have $\alpha(W\cup[x^k_i,a^k_i[_F)=\alpha(W)$ and
$\alpha(W\cup[x^l_j,a^l_j[_F)=\alpha(W)$. So by Claim \ref{claim*},
we obtain
\\ \hspace*{3cm} $\alpha(W\cup[x^k_i,a^k_i[_F\cup[x^l_j,a^l_j[_F)=\alpha(W)$.~~~~~~~~~~~~~~~~~~~~~~~~~~~~$(\star)$
\\ Also, taking the $[2,b]$-subgraph
$F'=(F-([x^k_i,a^k_i[_F\cup[x^l_j,a^l_j[_F))\cup Q\cup P_{kl}$, in
Claim \ref{claim1}, gives
$\alpha((W-Q)\cup[x^k_i,a^k_i[_F\cup[x^l_j,a^l_j[_F)>\alpha(W)$. But
as
$\alpha(W\cup[x^k_i,a^k_i[_F\cup[x^l_j,a^l_j[_F)\geq\alpha((W-Q)\cup[x^k_i,a^k_i[_F\cup[x^l_j,a^l_j[_F)$
hence we get
$\alpha(W\cup[x^k_i,a^k_i[_F\cup[x^l_j,a^l_j[_F)>\alpha(W)$ which
contradicts $(\star)$.$\Box$

\vspace{3mm} When $k=l$ in the previous claim, then we consider the structure of $D$.
If $D$ contains a cycle or $D$ is a tree
with two leaves $x_0$ and $y_0$ such that $N_F(x_0)=N_F(y_0)$, then the following claim gives the
path-independence of any couple of segments $P^{'k}_i$ and $P^{'k}_j$ in $\mathfrak{P'}$.

\begin{claim}\label{claim10}
Let $P^{'k}_i$ and $P^{'k}_j$ be two distinct segments of
$\mathfrak{P'}$. Suppose that $D$ contains a cycle or $D$ is a tree
with two leaves $x_0$ and $y_0$ such that $N_F(x_0)=N_F(y_0)$. Then
$P^{'k}_i$ and $P^{'k}_j$ are path-independent, for every $k, 1\leq
k\leq m'$.
\end{claim}

\emph{Proof of Claim \ref{claim10}.} Let $P^{'k}_i$ and $P^{'k}_j$
(with $1\leq i, j\leq m'$, $i\neq j$) be two segments in
$\mathfrak{P'}$. By way of contradiction, suppose that there is a
path $Q$ internally disjoint from $F\cup D$ joining a vertex
$a^k_i\in P^{'k}_i$ to a vertex $a^k_j\in P^{'k}_j$ and choose
$a^k_i$ and $a^k_j$ so that the sum
$d_F(x^k_i,a^k_i)+d_F(x^k_j,a^k_j)$ is minimum. \\ The segments
$[x^k_i,a^k_i[_F$ and $[x^k_j,a^k_j[_F$ verify the hypothesis of
Claim \ref{claim*}. Indeed, they are path-independent, by the choice
of $a^k_i$ and $a^k_j$. Furthermore, as
$[x^k_i,a^k_i[_F\subset[x^k_i,v^k_i[_F$ and
$[x^k_j,a^k_j[_F\subset[x^k_j,v^k_j[_F$ and by the choice of $v^k_i$
and $v^k_j$ we have $\alpha(W\cup[x^k_i,a^k_i[_F)=\alpha(W)$ and
$\alpha(W\cup[x^k_j,a^k_j[_F)=\alpha(W)$. So by Claim \ref{claim*},
we obtain
\\ \hspace*{3cm} $\alpha(W\cup[x^k_i,a^k_i[_F\cup[x^k_j,a^k_j[_F)=\alpha(W)$.~~~~~~~~~~~~~~~~~~~~~~~~~~~~$(\star\star)$
\\ On the other hand, if $D$ contains a cycle $C$, then let $Q'$ be a
path with internal vertices in $D-C$ joining $u_k$ to a vertex on
$C$. If $D$ is a tree with two leaves $x_0$ and $y_0$ such that
$N_F(x_0)=N_F(y_0)$. Then let $P$ be a path with internal vertices
in $D$ joining $x_0$ to $y_0$. Taking
$F'=(F-([x^k_i,a^k_i[_F\cup[x^k_j,a^k_j[_F))\cup Q\cup Q'\cup C$ in
the first case and $F'=(F-([x^k_i,a^k_i[_F\cup[x^k_j,a^k_j[_F))\cup
Q\cup u_kx_0Py_0u_k$ in the second one and using Claim \ref{claim1},
we obtain in both cases
$\alpha(W\cup[x^k_i,a^k_i[_F\cup[x^k_j,a^k_j[_F)\geq\alpha((W-Q)\cup[x^k_i,a^k_i[_F\cup[x^k_j,a^k_j[_F)>\alpha(W)$
which contradicts $(\star\star)$. $\Box$

\vspace{3mm} Suppose now that $D$ is either a trivial tree or $D$
has no leaves with the same neighborhood in $F$.

\begin{itemize}
\item If all couples of
distinct segments $(P^{'k}_i,P^{'k}_j)$ ($k, 1\leq k\leq m'$, $1\leq
i,j\leq|N'_F(u_k)|$) in $\mathfrak{P'}$ are path-independent then we
have finished. It is particularly the case if $s$ exists. Indeed, if
we suppose to the contrary that there exist two distinct segments
$P^{'k}_i$ and $P^{'k}_j$ ($k, 1\leq k\leq m'$, $1\leq
i,j\leq|N'_F(u_k)|$) in $\mathfrak{P'}$ that are path-dependent,
that is there is a path internally disjoint from $D\cup F$ joining a
vertex in $a^k_i\in P^{'k}_i$ to a vertex in $a^k_j\in P^{'k}_j$. We
choose these vertices so as to minimize the sum
$d_F(x^k_i,a^k_i)+d_F(x^k_j,a^k_j)$. Reasoning as in the previous
claims using Claim \ref{claim*} and taking in Claim \ref{claim1}
$F'=(F-[x^k_i,a^k_i[_F\cup[x^k_j,a^k_j[_F)\cup P_{kr}-u_rs$ where
$u_r$ is a neighbor of $s$ such that $r\neq k$, we get a
contradiction.
\\ It is also the case if there exists a vertex
$u_r$ $(1\leq r\leq m)$ that is of degree at most $b-1$ in $F$ or
that has in its neighborhood $N_F(u_r)$ a vertex $x$ such that $d_F(x)\geq3$.
Recall that in our case, this vertex is supposed to be put apart in
the procedure we have used. So, if we suppose that there is a path
internally disjoint from $D\cup F$ joining a vertex in $a^k_i\in
P^{'k}_i$ to a vertex in $a^k_j\in P^{'k}_j$ ($k\neq r$). We choose
these vertices so as to minimize the sum
$d_F(x^k_i,a^k_i)+d_F(x^k_j,a^k_j)$. Here again, using Claim
\ref{claim*} and taking in Claim \ref{claim1},
$F'=(F-[x^k_i,a^k_i[_F\cup[x^k_j,a^k_j[_F)\cup P_{kr}$ if
$d_F(u_r)\leq b-1$ or $F'=(F-[x^k_i,a^k_i[_F\cup[x^k_j,a^k_j[_F)\cup
P_{kr}-u_rx$ if $d_F(u_r)=b$ and $d_F(x)\geq3$ where $x\in
N'_F(u_r)$, we get a contradiction.

\item If not, then this case is treated in following claim.
\end{itemize}

\begin{claim}\label{claim11}
Suppose that $D$ is a trivial tree or a tree with no leaves having
the same neighborhood in $F$. Suppose moreover that there exist two
distinct segments $P^{'k}_i$ and $P^{'k}_j$ ($k$, $1\leq k\leq m'$,
$1\leq i,j\leq|N'_F(u_k)|$) in $\mathfrak{P'}$ that are
path-dependent. Then there exists no other couple of segments
$(P^{'l}_p,P^{'l}_q)$ ($l\neq k$, $1\leq l\leq m'$, $1\leq
p,q\leq|N'_F(u_l)|$, $p\neq q$) in $\mathfrak{P'}$ that are
path-dependent.
\end{claim}

\emph{Proof of Claim \ref{claim11}.} The proof is basically the same
as the previous. Let $a_i^k$ and $a_j^k$ be two vertices in
$P^{'k}_i$ and $P^{'k}_j$ respectively that are joined by a path
internally disjoint from $F\cup D$ and chosen so as to minimize the
sum $d_F(x^k_i,a^k_i)+d_F(x^k_j,a^k_j)$. Suppose to the contrary
that there exist two distinct segments $P^{'l}_p,P^{'l}_q$ ($l\neq k$, $1\leq
l\leq m'$, $1\leq p,q\leq|N'_F(u_l)|$) and a path internally
disjoint from $F\cup D$ joining a vertex $a_p^l\in P^{'l}_p$ to a
vertex $a_q^l\in P^{'l}_q$ and choose these vertices in such a way
that $d_F(x^l_p,a^l_p)+d_F(x^l_q,a^l_q)$ is minimum. Then using
Claim \ref{claim*} with four intervals and taking
$F'=(F-([x^k_i,a^k_i[_F\cup[x^k_j,a^k_j[_F)-([x^l_p,a^l_p[_F\cup[x^l_q,a^l_q[_F))\cup
P_{kl}$ in Claim \ref{claim1} yields a contradiction. As $k\neq l$,
then Claim \ref{claim9} guarantees the path-independence of the
segments $P^{'k}_r,P^{'l}_t$ for $r\in\{i,j\} , t\in\{p,q\}$. $\Box$

By the claims above, we have that $\mathfrak{P'}$ contains
several segments that are path-independent.
Furthermore, the following remark claims that an additional segment can be considered
when needed, particularly when $S$ contains a vertex of degree 3.

\begin{remark}\label{rmk2}
If there exists a vertex $s\in S$ such that $d_F(s)=3$ ($s$ is
unique by Claim \ref{claim+}(2)). We consider two cases:
\\(i) If $s$ is in the neighborhood of three vertices $u_k$, $u_l$ and $u_p$, with $1\leq k, l, p \leq m$ and $k, l, p$ pairwise distinct.
Then setting $P^*=\{s\}$ we have that $P^*$ is path-independent from
any path in $\mathfrak{P'}$ and $\alpha(W\cup
P^*)>\alpha(W)$.
\\(ii) If $s$ is in the neighborhood of exactly two vertices, say $u_k$ and $u_l$,
$k\neq l, 1\leq l, k\leq m$. If furthermore $\mathfrak{P}$ does not
contain paths of Type 3, then there exists a path $P^*$ that is
path-independent from any segment in $\mathfrak{P'}$ included in a
path of Type 1 or Type 2. Furthermore $\alpha(W\cup P^*)>\alpha(W)$.
\end{remark}

\emph{Proof.}
\\(i) First, taking $F'=(F-\{s\})\cup P_{kl}$ in Claim \ref{claim1}
we obtain $\alpha(W\cup\{s\})>\alpha(W)$. Of course, since by Claim
\ref{claim+}(2) $d_F(u_i)=b$ for all $i=1,...,m$, then we have that
$F'$ is a $[2,b]$-subgarph of $G$. As $s\in
N_F(u_k)\cap N_F(u_l)\cap N_F(u_p)\subset N_F(u_k)\cap N_F(u_l)$
then we can write $\{s\}=P^{k}_i$ or $\{s\}=P^{l}_i$ as suitable to apply Claim \ref{claim9}
and show the path-independence of $\{s\}$ from any segment in $\mathfrak{P'}$.
\\(ii) Let us start from $s$ and go forward following the chosen
orientation from a vertex of degree 2 in $F$ to a vertex of degree 2
in $F$, until coming across a vertex $y$ whose successor $y^+$ is of
degree at least 3. We have that $y^+\notin\{u_1,\ldots,u_m\}$,
otherwise, going in the opposite direction, the segment $[y,s[_F$ (where $s$ is not taken)
is a path of Type 3. Moreover, $y^+\notin\cup_{i=1}^m
N_F(u_i)$ because since $s$ exists then by Claim \ref{claim+} every
vertex in $\cup_{i=1}^m N_F(u_i)$ is of degree 2. So $y^+\in
V(F)-\cup_{i=1}^m N_F[u_i]$. The path $P=s\ldots y$ can be
considered as a path deriving from $u_k$ ($P=P^{'k}_i$) or deriving
from $u_l$ ($P=P^{'l}_i$) and hence reasoning as in Claim
\ref{claim8}, taking $F'=F-P$, we obtain $\alpha(W\cup
P)>\alpha(W)$. Let $v$ be the first vertex of $P$ following the
chosen orientation such that $\alpha(W\cup[s,v]_F)>\alpha(W)$.
Setting $P^*=[s,v]_F$, we can show its path-independence with any
segment in $\mathfrak{P'}$ included in a path of Type 1 or Type 2,
like in Claim\ref{claim9}. $\Box$

\vspace{3mm}

Finally, to count the number of pairwise path-independent segments
in $\mathfrak{P'}$, those whose independence is guaranteed by Claims
\ref{claim9}, \ref{claim10}, and \ref{claim11}, we distinguish
different cases according to the structure of $D$ and get in any
case, at least $\lfloor\frac{b(\delta-1)}{2}\rfloor$ (recall that
$m'\geq\delta-1$) path-independent segments, adding when necessary
the path $P^*$ (in particular when $s$ exists).
Notice that when $\mathfrak{P_1}$
contains paths of Type 3, then in these paths one vertex in $\cup_{i=1}^m N_F(u_i)$
is used at once, so the bound $\lfloor\frac{b(\delta-1)}{2}\rfloor$ holds,
otherwise $P^*$ is added.

The segments in $\mathfrak{P'}\cup\{P^*\}$, when added to $W\cup D$
augment $\alpha(W\cup D)$. Put
$\mathfrak{L}=\mathfrak{P'}\cup\{P^*\}$. Recall that the segments of
$\mathfrak{L}$ are independent from $D$ by construction. For each
$P\in\mathfrak{L}$, let $W_P$ be the union of components of $W$ that
contain a neighbor of $P$.

We have that
\\ $\alpha(W\bigcup\cup_{P\in\mathfrak{L}}P)=\alpha(W-\cup_{P\in\mathfrak{L}}W_{P})
+\sum_{P\in\mathfrak{L}}\alpha(W_{P}\cup P)$
\\ \hspace*{2.9cm}$\geq\alpha(W-\cup_{P\in\mathfrak{L}}W_{P})
+\sum_{P\in\mathfrak{L}}\alpha(W_{P})+|\mathfrak{L}|$
\\
\hspace*{2.9cm}$\geq\alpha(W)+\lfloor\frac{b}{2}(\delta-1)\rfloor$.
\\Hence
\\ $\alpha=\alpha(G)=\alpha(W\cup D\cup F)\geq\alpha(W\cup
D\bigcup\cup_{P\in\mathfrak{L}}P)$
\\ \hspace*{5.1cm}$\geq\alpha(D)+\alpha(W\bigcup\cup_{P\in\mathfrak{L}}P)$
\\ \hspace*{5.1cm}$\geq\alpha(D)+\alpha(W)+\lfloor\frac{b}{2}(\delta-1)\rfloor$
\\ \hspace*{5.1cm}$=\alpha(W\cup
D)+\lfloor\frac{b}{2}(\delta-1)\rfloor$.
\\ So $\alpha(W\cup D)=\alpha(G-F)\leq\alpha-\lfloor\frac{b}{2}(\delta-1)\rfloor$
and the proof of Theorem \ref{thm2} is achieved. $\blacksquare$

\vspace{3mm}

\textbf{Proof of Theorem \ref{thm1}.} Since by Theorem \ref{thm2},
$\alpha(G-F)\leq\alpha-\lfloor\frac{b}{2}(\delta-1)\rfloor$, then
the subgraph of $G$ induced by $V(W\cup D)=V(G-F)$ can be covered by
at most $\alpha-\lfloor\frac{b}{2}(\delta-1)\rfloor$ cycles, edges
or vertices (see for instance \cite{posa}). Denote by $\mathcal{E}$
the set of cycles, edges or vertices covering $G-F$. The graph
$F\cup \mathcal{E}$ is a pseudo $[2,b]$-factor of $G$ with at most
$\alpha-\lfloor\frac{b}{2}(\delta-1)\rfloor$ edges or vertices. This
completes the proof of Theorem \ref{thm1}. $\blacksquare$

\end{document}